\pdfoutput=1 

\PassOptionsToPackage{unicode}{hyperref}
\PassOptionsToPackage{hyphens}{url}
\PassOptionsToPackage{dvipsnames,svgnames,x11names}{xcolor}

\documentclass[12pt]{article}

\usepackage{amsmath,amssymb}
\usepackage[T1]{fontenc}
\usepackage[utf8]{inputenc}
\usepackage{textcomp}
\usepackage{lmodern}
\usepackage{graphicx}
\usepackage{orcidlink}

\IfFileExists{upquote.sty}{\usepackage{upquote}}{}
\IfFileExists{microtype.sty}{%
	\usepackage[]{microtype}
	\UseMicrotypeSet[protrusion]{basicmath} 
}{}

\IfFileExists{xurl.sty}{\usepackage{xurl}}{} 
\usepackage[unicode]{hyperref}
\usepackage{bookmark}

\hypersetup{
	pdftitle   = {Title},
	pdfauthor  = {Author 1; Author 2},
	pdfkeywords= {3 to 6 keywords, that do not appear in the title},
	colorlinks = false,     
	hidelinks  = true,      
	linkcolor  = blue,
	filecolor  = Maroon,
	citecolor  = Blue,
	urlcolor   = Blue,
	breaklinks = true,
	pdfcreator = {LaTeX via pandoc}
}

\usepackage{natbib}
\newcommand{\anon}{1}

\begin{document}

\def\spacingset#1{\renewcommand{\baselinestretch}%
{#1}\small\normalsize} \spacingset{1}


\if1\anon
{
	\title{\bf Factors Associated with Unit-Specific Failure in a University-Level Statistics Course}
	\author{Biviana Marcela Suárez Sierra\,\orcidlink{0000-0003-2151-3537}
		\thanks{
			I gratefully acknowledge the students enrolled in the General Statistics course during the 2024-2 semester for their participation in the survey and their constructive feedback. I also thank the instructors who contributed to both lectures and laboratory sessions, and Professor Nicolás Moreno for the several enriching discussions about this work. \textit{No funding was received for this study.}}\hspace{.2cm}\\
		School of Applied Sciences and Engineering, EAFIT University\\
	}
	\maketitle
} \fi

\if0\anon
{
	\bigskip
	\bigskip
	\bigskip
	\begin{center}
		{\LARGE\bf Factors Associated with Unit-Specific Failure in a University-Level Statistics Course}
	\end{center}
	\medskip
} \fi

\bigskip
\begin{abstract}
This study investigates the factors associated with failure in each of the four thematic units of a General Statistics course offered at a private university in Colombia. Unlike traditional analyses that treat performance as a single outcome, this research disaggregates results by unit—Exploratory Data Analysis, Probability and Random Variables, Statistical Inference, and Linear Regression—highlighting distinct challenges across content areas. Based on a sample of 186 undergraduate students from Engineering, Geology, and Interactive Design programs, the study combines exam performance data with self-perceived preparedness surveys to develop unit-specific logistic regression models. The findings reveal consistent structural disadvantages for students from non-engineering programs, especially in concept-heavy units such as Inference and Regression. Academic stage and perception of competence also emerged as important predictors, though their effects varied across units. The results align with prior research on statistical thinking and self-efficacy, and support the need for targeted pedagogical interventions and curricular alignment. This disaggregated approach offers a more nuanced understanding of academic vulnerability in statistics education and contributes to the design of evidence-based, context-sensitive strategies to reduce failure and improve learning outcomes.
\end{abstract}

\noindent%
\vfill

\newpage
\spacingset{1.8} 

\section{Introduction}\label{sec-intro}

Statistics is widely recognized as a foundational course across undergraduate curricula, particularly in science, technology, engineering, and mathematics (STEM) disciplines. It plays a critical role in developing students’ capacity for analytical thinking, data-driven reasoning, and evidence-based decision-making. Despite its importance, introductory statistics remains one of the most challenging subjects for students, often marked by low engagement, high dropout rates, and widespread conceptual difficulties. These issues are especially acute in multidisciplinary contexts, where students arrive with heterogeneous backgrounds in mathematics, computing, and scientific reasoning.

While numerous studies have examined general factors affecting performance in statistics education—such as prior preparation, instructor practices, or students’ attitudes—there is a noticeable gap in the literature concerning performance variation within the course itself. Most existing research focuses on aggregate course outcomes rather than exploring how performance and failure rates may vary across individual content units. This lack of disaggregation limits our understanding of which topics pose the greatest barriers to learning and hinders the design of targeted pedagogical interventions.

To address this gap, the present study investigates the unit-specific factors associated with failure rates in a General Statistics course taught at a private university in Colombia. The course is required for students in Engineering, Geology, and Interactive Design programs and is structured around four key units: (1) Exploratory Data Analysis, (2) Probability Laws, Bayes’ Theorem, and Random Variables, (3) Statistical Inference, and (4) Linear Regression. By analyzing both academic outcomes and students’ self-perceived preparedness for each unit, this study seeks to uncover which pedagogical, cognitive, or contextual factors are most predictive of unit-level success or failure.
The following research questions guide the analysis:
\begin{itemize}
	\item How do failure rates vary across the four course units?
	\item What pedagogical, cognitive, or contextual factors explain these variations?
\end{itemize}
By focusing on disaggregated unit performance, this study contributes to a more nuanced understanding of academic vulnerability in statistics education. The findings are intended to inform course redesign efforts, promote equity across disciplines, and enhance student learning through evidence-based, unit-specific interventions.

\section{Theoretical Framework}\label{sec-framework}

Understanding the factors that influence student performance in introductory statistics courses has long been a central concern in educational research, particularly in STEM fields. Recent efforts have aimed to improve instructional effectiveness and reduce failure rates by identifying institutional, cognitive, and affective variables that shape learning outcomes. \citet{burns2020assessing}, for instance, developed a statistical model to evaluate performance in an engineering statistics course using average exam scores. Their results emphasized the relevance of objective indicators—such as homework and lab grades—and contextual variables like academic program, class section, and teaching assistant assignments. Interestingly, attitudinal responses and engagement metrics showed no significant relationship with performance, suggesting that self-reported sentiment alone may have limited predictive power when not complemented by behavioral or cognitive measures.

Motivated by this work, the present study explores the predictors of failure across four thematic units of a General Statistics course taught at a private university. While the course structure differs from that of engineering education, we adopt a similar multivariable framework incorporating academic records, instructional characteristics, and students’ self-perceptions. Unlike \citet{burns2020assessing}, our context excludes homework scores and engagement metrics, but includes self-perceived preparedness indicators captured before each exam.

Our study setting features several structural supports: peer tutoring offered by high-achieving students, laboratory sessions led by adjunct instructors, and differentiated teaching across three groups. Exams are administered synchronously under faculty supervision, and each includes a short perception survey on the unit’s learning outcomes. However, key factors remain beyond experimental control, such as students’ attendance (not mandatory), their disciplinary background, and their academic trajectory.

Building on work by \citet{garfield2007how} and \cite{garfield2008developing}, we also consider conceptual challenges associated with statistical thinking. Their review emphasizes that foundational concepts such as the mean and variability remain difficult to grasp even among high-performing students. Many learners rely on procedural recall rather than genuine conceptual understanding. Moreover, students often perceive data sets as collections of isolated values rather than as aggregates with emergent properties—representing a significant cognitive hurdle in developing statistical reasoning.

Understanding variability, for instance, is frequently reduced to calculating standard deviation, range, or interquartile range, without interpreting their role in broader inference tasks. As \citet{garfield2007how}, \cite{garfield2008developing} argue, this stems not only from conceptual complexity but also from intuitive misconceptions that contradict statistical logic. Learning is most effective when students are encouraged to predict outcomes, confront cognitive conflicts, and revise their interpretations in light of new evidence.

To address these challenges, the literature recommends connecting randomness with data analysis to foster statistical reasoning—defined as the ability to integrate concepts related to uncertainty and data. Teaching should aim at developing literacy, reasoning, and thinking in a hierarchical manner, avoiding fragmented instruction. Conceptual integration is key to mastering statistical ideas such as distribution, sampling, and inference, which cannot be taught in isolation \citep{garfield2007how} and \cite{garfield2008developing}.

Another critical dimension involves students’ attitudes toward statistics. Several studies have shown that these attitudes are often resistant to change, and in some cases worsen over time \citep{garfield2007how}. While some students complete courses with strong grades, they retain negative views about the subject’s difficulty and relevance. Nevertheless, \citet{finney2003self} and \citet{bandura1997self} highlight that motivation, perseverance, and a desire to learn are more predictive of academic success than prior knowledge or initial attitudes. Our own data reflect this pattern: students with limited math backgrounds sometimes outperformed peers when highly engaged and motivated.

The persistent gap between perceived and actual utility of statistics is also documented by \citet{ghulami2015attitudes}, \cite{Fernandez2025} and \citet{gal1994beliefs}, who describe students’ emotional detachment and frustration despite recognizing the long-term importance of the discipline. They call for a “rebranding” of statistics instruction—emphasizing collaborative learning, authentic experiences, and a reduced focus on manual calculation. These insights reinforce the need for pedagogical strategies that humanize the subject and make its applications visible in real-world contexts.

To quantitatively assess students’ attitudes, the SATS-36 scale remains one of the most validated instruments, measuring dimensions such as \textit{Affect}, \textit{Cognitive Competence}, \textit{Value}, \textit{Difficulty}, \textit{Interest}, and \textit{Effort}. Its psychometric structure has been confirmed in diverse contexts and continues to evolve, including through Rasch model analyses for cross-cultural adaptability \citep{rasch2023sats}.

Finally, efforts to modernize statistics education, such as those described by \citet{burns2019redesigning}, provide relevant models. Their redesigned multidisciplinary course shares similarities with our own institutional context, including diverse student populations and a lecture-lab format. However, whereas their approach de-emphasizes software in assessments, our model places computational thinking—centered around \texttt{R}—at the core of evaluations. Both settings face common challenges: low engagement, preference for procedural learning, and resistance to abstract reasoning. Yet, the implementation of semester-long projects, as proposed in their work, remains an area of opportunity for our context.

In sum, the theoretical framework for this study draws on interdisciplinary literature on performance predictors, cognitive development, and student affect. It situates our investigation within broader debates on how best to teach statistics—acknowledging structural, perceptual, and motivational variables that influence failure rates across different content units.

\section{Methods}\label{sec-meth}

	This study employed a cross-sectional quantitative design, complemented by student surveys. The quantitative component was based on institutional data derived from four digitally administered, real-time partial exams---each aligned with a specific thematic unit of the General Statistics course. These exams, conducted under the supervision of the course instructor, served as the main performance metric. Complementary to this, Likert-scale surveys were administered to students prior to each exam, capturing their self-perceived understanding of the unit content, level of preparation, and confidence in facing the upcoming evaluation.
	
	Each unit in the course was aligned with a specific learning outcome. Unit 1, ``Descriptive Statistics of a Variable,'' assessed students' ability to summarize real-world data using tables, graphs, and summary statistics. Unit 2, ``Analysis of Random Phenomena,'' focused on understanding probability and random variables through experimental contexts. Unit 3, ``Statistical Inference,'' evaluated students’ ability to infer population parameters through estimation and hypothesis testing. Unit 4, ``Analysis of Linear Dependence,'' assessed students' capacity to interpret and estimate parameters in linear regression models. By integrating performance data with pre-assessment self-perceptions, the study enables a meaningful comparison between students’ perceived understanding and their actual performance, offering a more comprehensive picture of the learning process.
	
	The sample comprised 186 students enrolled in the General Statistics course during the 2024-2 academic term. The course is offered by the School of Applied Sciences and Engineering (ECAeI by its initials in Spanish) and is also required for students in the Interactive Design program, part of the School of Arts and Humanities. Due to differences in curricular structure, students take the course at varying stages in their undergraduate careers. Nevertheless, 74\% were enrolled between the fourth and sixth semesters, while 58\% were in the fifth semester or earlier. Of the total sample, 24\% belonged to the Interactive Design program, and only 9\% were in the group led by the full-time faculty member. In one survey task, students were asked to write the name of the department where they completed secondary school following specific instructions; only 52\% followed the instructions correctly. Additionally, 69\% reported prior exposure to statistics in high school, 6\% indicated no prior programming experience, and 44\% reported receiving some form of financial aid.
	
	The dependent variable was the failure rate for each unit, defined as a binary outcome (pass/fail) based on exam performance. The observed failure rates were 46\% for Unit 1, 19\% for Unit 2, 40\% for Unit 3, and 30\% for Unit 4. The independent variables considered included the instructional approach associated with each teaching group. Variables such as prior academic performance, attendance, and tutoring participation were excluded due to the absence of standardized records for the term analyzed.
	
	The partial exam results allowed for unit-specific analysis, revealing substantial gaps in understanding of key statistical concepts. In Unit 1, only 36.02\% of students correctly solved the tabular data representation task. Even more striking, only 23.66\% succeeded in items assessing the combined application of central tendency, position, and dispersion measures. These results are consistent with findings by \citet{garfield2007how}, which highlight that seemingly basic concepts such as the mean and variability remain difficult to master---even for high-achieving students.
	
	In Unit 3, centered on statistical inference, the learning gap was even wider. Only 39.25\% of students mastered problems involving the sampling distribution of the variance. Just 21.51\% correctly answered general confidence interval items, and only 43.55\% did so for intervals specifically related to variance. The most critical result was in hypothesis testing: only 6.45\% accurately solved a comprehensive exercise. Disaggregated data showed that 30.65\% answered hypothesis testing questions about the mean correctly, 47.85\% about proportions, and 31.72\% about variance. These results underscore the study’s broader conclusion: procedural competence does not equate to conceptual understanding, reinforcing the need for learning strategies that emphasize statistical reasoning and critical analysis.
	
	In Unit 4, which addressed linear regression, 47.31\% of students excelled in solving problems involving linear regression and interpreting the correlation coefficient. While this figure was higher than in previous units, it still suggests the need to strengthen instruction that prioritizes interpretation and application, particularly in real-world contexts. Taken together, these findings suggest that while some students develop strong analytical skills, many continue to struggle with meeting key learning objectives of the course.
	
	To investigate the predictors of failure, binary logistic regression models were developed for each unit, using failure (1 = failed, 0 = passed) as the dependent variable.
	
	For Unit 1, the model included quantitative variables such as age (based on year of birth), exam duration (dur1), and three self-perception variables---Perp1\_P1 (ability to recognize the nature of data), Perp2\_P1 (ability to summarize data using tables), and Perp3\_P1 (ability to create visualizations and calculate and interpret statistical measures of position, central tendency, and dispersion). These were originally Likert-scaled and transformed to a 0--1 range. Categorical dummy variables included whether the student graduated from a public school, lacked programming experience, was assigned male at birth, followed instructions correctly, was enrolled in $\leq$ fifth semester (etapa\_sem), belonged to the Interactive Design program, had prior experience in statistics, completed high school outside Antioquia, and was assigned to the full-time faculty member's group. Model selection was based on deviance tests comparing the full model to reduced alternatives; the full model demonstrated better fit.
	
	Although the survey included open-ended items (e.g., birth year, school name, and graduation department), these were not used for qualitative analysis, as they were not designed to capture interpretive responses.
	
	The model for Unit 2 followed a similar structure. The dependent variable, Reprobation\_P2, captured failure on the second exam. Predictors included age, duration (dur2), and five perception variables: Perp1\_P2 (understanding of continuous distributions), Perp2\_P2 (discrete distributions), Perp3\_P2 (probability properties), Perp4\_P2 (Bayes' theorem and total probability), and Perp5\_P2 (random variable properties). Dummy variables mirrored those in Unit 1.
	
	In Unit 3, which focused on inference, the model included the following self-perception variables: Perp1\_P3 (recognition of key statistics and their distributions), Perp2\_P3 (conditions for applying the Central Limit Theorem), Perp3\_P3 (evaluation of point estimators), Perp4\_P3 (confidence intervals), and Perp5\_P3 (hypothesis testing using rejection regions and p-values). Additionally, a categorical variable was included to capture the student’s academic stage (etapa\_sem), coded as 1 if the student was in the early semesters of their degree program. The full model outperformed all reduced alternatives.
	
	The Unit 4 model, centered on regression, included three perception variables: Perp1\_P4 (identifying linear relationships and computing coefficients), Perp2\_P4 (testing regression coefficients and interpreting results), and Perp3\_P4 (interpreting $R^2$ and standard error of the estimate). The same dummy variables were included, along with academic stage. As with previous units, the full model showed the best fit and was retained for the analysis and interpretation of results.

\section{Results by Thematic Unit}\label{sec-results}

\subsection{Unit 1: Descriptive Statistics of a Variable}

A binary logistic regression model was estimated to identify the factors associated with failure in the first partial exam, which corresponds to the unit on descriptive statistics of a single variable. The dependent variable was failure (1 = fail, 0 = pass), and the model included sociodemographic, academic, and self-perception variables.

Among the fourteen predictors, four showed statistically or substantively relevant effects. The variable \texttt{No\_ECAeI} (students from programs outside the School of Engineering) had a positive and highly significant effect on the likelihood of failure ($\beta = 1.226$, $p = 0.00284$), with an odds ratio (OR) of $e^{1.226} \approx 3.41$. This suggests that students not enrolled in engineering-related programs were 3.41 times more likely to fail, possibly due to differences in prior preparation or alignment with the course's conceptual focus.

The variable \texttt{Grupo\_B}, representing students assigned to the section led by the full-time faculty member, was also significant ($\beta = 1.222$, $p = 0.0476$, OR $\approx 2.28$). Students in this group had more than twice the odds of failing compared to those in other sections. This may be due to stricter evaluation criteria, less flexibility during exams, or differences in instructional style.

Two variables were marginally significant at the 10\% level: prior experience in statistics (\texttt{eee}) ($\beta = -0.624$, $p = 0.0845$) and the perception variable \texttt{Perp1\_P1} (self-assessed ability to recognize the nature of data) ($\beta = -2.694$, $p = 0.0622$). For each 0.1-point increase in \texttt{Perp1\_P1}, the odds of failure decreased by 23.6\% ($OR = e^{-0.269} \approx 0.764$), highlighting the role of self-confidence in specific competencies.

Other variables such as age, sex, school type, programming experience, and test duration showed no significant effect. The model showed a notable improvement over the null model (AIC = 255.35, residual deviance = 225.35), suggesting reasonable fit.

\subsection{Unit 2: Analysis of Random Phenomena}

The model for the second unit, focused on probability and random variables, identified six relevant predictors—four significant at the 1\% level, one at 5\%, and one marginally significant at 10\%.

Age showed a significant positive effect ($\beta = 0.230$, $p = 0.008$), with an OR of 1.26, implying that each additional year of age increased the odds of failing by 25.8\%. Students without programming experience (\texttt{exP}) had OR $\approx 6.42$ ($\beta = 1.859$, $p = 0.018$), highlighting the protective role of logical and computational reasoning skills.

The sex variable (\texttt{sex}) showed a significant negative effect ($\beta = -1.061$, $p = 0.035$, OR $\approx 0.35$), indicating that male students had 65\% lower odds of failing than females, all else equal—an observation that warrants deeper investigation into gender dynamics in technical education.

The variable \texttt{No\_ECAeI} remained significant ($\beta = 1.152$, $p = 0.030$, OR $\approx 3.16$), confirming the disadvantage faced by students from non-engineering backgrounds. A marginally significant perception variable \texttt{Perp5\_P2} ($\beta = -3.753$, $p = 0.059$) showed that improved self-perception in random variable topics reduced failure odds by 31.3\% per 0.1-point increase.

The model’s residual deviance was 142.38 (AIC = 176.38), suggesting strong explanatory power. Collectively, these results emphasize the impact of both structural (discipline, programming background) and individual (age, gender, self-perception) factors in statistical reasoning assessments.

\subsection{Unit 3: Statistical Inference}

In the third unit on statistical inference, two variables were relevant: \texttt{No\_ECAeI} ($\beta = 0.841$, $p = 0.0468$, OR $\approx 2.32$) and \texttt{etapa\_sem} (being in the early semesters) ($\beta = 0.948$, $p = 0.0737$, OR $\approx 2.58$).

Students from non-engineering programs were more than twice as likely to fail, while those in earlier academic stages had 2.58 times higher odds of failure. These results suggest the importance of prior academic exposure to abstract reasoning and inferential thinking.

Interestingly, the effect of \texttt{Grupo\_B} disappeared in this model ($p = 0.802$), possibly due to students adapting to the teaching style or converging evaluation strategies across sections. The model’s AIC was 264.44 with a residual deviance of 228.44.

\subsection{Unit 4: Linear Dependence Analysis}

The fourth unit addressed regression and correlation. The model included three significant predictors and two marginally significant ones.

The perception variable \texttt{Perp2\_P4} (confidence in hypothesis testing on regression coefficients) had a strong protective effect ($\beta = -6.524$, $p = 0.008$, OR $\approx 0.0015$). Even a 0.1-point increase in this perception lowered the odds of failing by 48\%, highlighting the crucial role of conceptual mastery.

Test duration (\texttt{dur4}) was also significant ($\beta = -0.030$, $p = 0.0468$, OR $\approx 0.971$), suggesting a 2.9\% decrease in failure odds for each extra minute spent.

\texttt{No\_ECAeI} again appeared significant ($\beta = 1.093$, $p = 0.026$, OR $\approx 2.98$), confirming the consistent disadvantage for students outside engineering programs.

Two marginal effects were observed: \texttt{ind} ($\beta = 0.713$, $p = 0.0688$, OR $\approx 2.04$) and \texttt{Fuera\_Antioquia} ($\beta = -0.831$, $p = 0.0908$, OR $\approx 0.436$). The first, surprisingly, suggested higher failure risk for those who reported following instructions, potentially due to coding issues. The latter suggested that students who completed high school outside Antioquia had 56\% lower odds of failure, possibly reflecting educational context differences.

The model’s AIC was 221.12, and the residual deviance was 189.12, indicating a solid model fit.

\subsection{Comparative Reflections Across Units}

A consistent finding across all four units was the significant disadvantage associated with \texttt{No\_ECAeI}. This suggests a structural gap in preparedness or curricular alignment between engineering and non-engineering students, particularly those from arts or design programs.

Self-perception variables were particularly important in Units 1 and 4, reinforcing the idea that confidence in core competencies is a strong predictor of success, especially in conceptual and inferential topics.

The influence of the instructor group (\texttt{Grupo\_B}) was significant only in Unit 1; for the remaining units, this variable was no longer significant, likely due to adaptation or the homogenization of assessment conditions.

Factors like age, gender, programming experience, and academic progression showed varying effects across units, indicating that student vulnerability is context-specific and must be understood in relation to the content and cognitive demands of each unit.

Overall, the results underscore the multifactorial nature of academic performance in statistics, suggesting that effective pedagogical strategies should combine content-level interventions with structural and individualized academic support.

\subsection{Graphical Analysis of Failure Risk by Sociodemographic, Academic, and Self-Perception Factors}

\begin{figure}[h]
	\centering
	\includegraphics{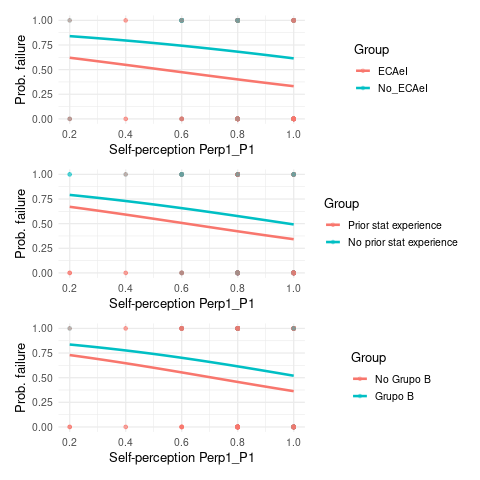}
	\caption{Estimated probability of failure on Exam 1 as a function of self-perceived ability to recognize the nature of data (Perp1\_P1), stratified by key categorical variables. The top panel compares students from ECAeI and non-ECAeI programs; the middle panel contrasts students with and without prior experience in statistics; the bottom panel differentiates students assigned to Group B (full-time instructor) versus other groups. The graphs reveal that higher self-perception is consistently associated with lower failure risk, although the magnitude of the effect varies by subgroup.}
	\label{fig:Fig1}
\end{figure}

Figure~\ref{fig:Fig1} displays the estimated probability of failing the first exam as a function of students’ self-perceived ability to recognize the nature of data (Perp1\_P1), stratified by key categorical variables. A consistent inverse relationship emerges across all panels: higher levels of perceived competence are associated with lower failure risk. However, the magnitude and baseline probabilities vary across subgroups. In the top panel, students from non-ECAeI programs consistently exhibit higher failure probabilities across the perception scale, suggesting a structural disadvantage tied to academic background. The middle panel reveals that students without prior exposure to statistics face significantly higher failure risks, especially at lower levels of self-perception, underscoring the protective role of previous experience. The bottom panel compares students taught by the full-time instructor (Group B) to those in other sections. While the downward trend remains consistent, Group B students show higher baseline failure probabilities throughout. These patterns highlight the need to consider both cognitive and contextual dimensions when assessing academic risk in the descriptive statistics unit.

\begin{figure}[h]
	\centering
	\includegraphics{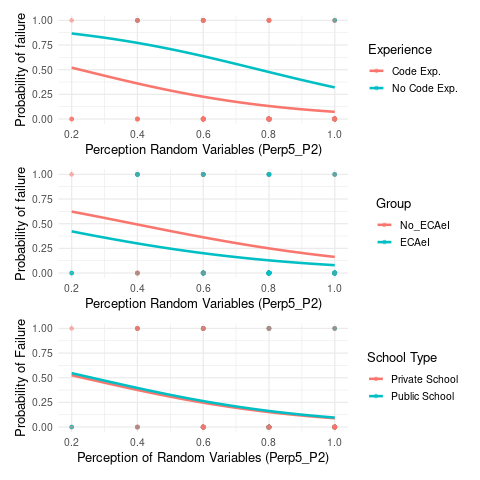}
	\caption{Predicted probability of failing the second midterm exam as a function of students’ self-perceived ability to work with random variables (Perp5\_P2), disaggregated by (top) prior coding experience, (middle) academic program (Interactive Design vs. ECAeI), and (bottom) type of high school (public vs. private). Each line represents a logistic regression prediction, controlling for other variables in the model. A consistent negative association is observed between perception and failure risk, with baseline differences across groups.}
	\label{fig:Fig2}
\end{figure}

Figure~\ref{fig:Fig2} explores failure probabilities in the second midterm, focusing on students’ perceived ability to work with random variables (Perp5\_P2). The plots, disaggregated by prior programming experience, academic program, and type of high school, reveal a consistent negative association: as perceived competence increases, failure risk decreases. Yet, baseline differences persist. Students without coding experience show higher failure probabilities across the scale (top panel), suggesting that computational familiarity may mitigate difficulties in probability-related topics. In the middle panel, students from non-engineering programs again display elevated risk—even at higher perception levels—while the bottom panel shows only modest effects by high school type, with public school students slightly more vulnerable across perception levels. Overall, Figure~\ref{fig:Fig2} underscores the interplay between individual skills and academic context in shaping performance.

\begin{figure}[h]
	\centering
	\includegraphics{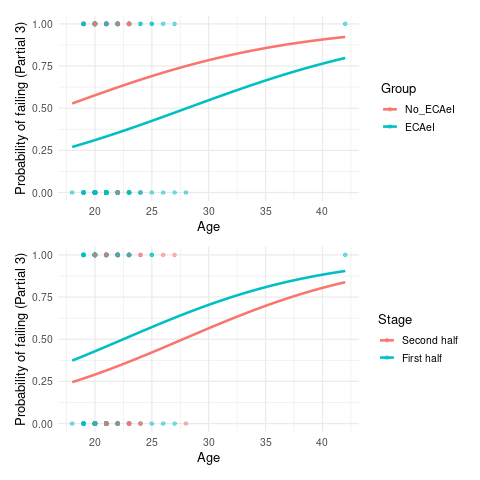}
	\caption{Estimated probability of failing the third partial exam as a function of age, stratified by academic program (top panel: ECAeI vs. non-ECAeI students) and academic stage (bottom panel: first vs. second half of undergraduate studies). In both cases, the probability of failure increases with age, but is consistently higher for students outside ECAeI and for those in the early stages of their academic programs.}
	\label{fig:Fig3}
\end{figure}

Figure~\ref{fig:Fig3} examines the relationship between age and failure on the third exam, focusing on academic program (top panel) and academic stage (bottom panel). In both, failure probability increases with age, although the trend varies across groups. Non-ECAeI students consistently face higher failure risk at all ages, reinforcing earlier findings of structural academic disadvantage. Similarly, students in the early stages of their undergraduate programs are more likely to fail than their advanced peers, regardless of age. This suggests that both maturity and academic progression act as protective factors, particularly in conceptually demanding units like statistical inference. Additionally, age may reflect more complex trajectories, such as interrupted studies or delayed entry, which could further influence performance.

\begin{figure}[h]
	\centering
	\includegraphics[width=0.9\linewidth]{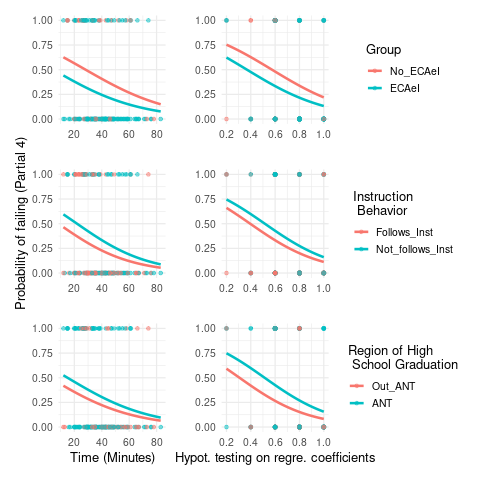}
	\caption{Estimated probability of failing the fourth partial exam (Unit 4) as a function of exam duration (in minutes) and the perception variable \texttt{Perp2\_P4} (confidence in hypothesis testing on regression coefficients), stratified by three factors: academic program (ECAeI vs. non-ECAeI), instructional compliance behavior (follows vs. does not follow instructions), and high school graduation region (Antioquia vs. outside Antioquia).}
	\label{fig:partial4-pred}
\end{figure}

Finally, Figure~\ref{fig:partial4-pred} presents the estimated probability of failing the fourth exam (Unit 4), based on exam duration, the perception variable \texttt{Perp2\_P4} (confidence in hypothesis testing on regression coefficients), and academic background. Across all panels, longer test duration and higher perceived competence in this area correlate with lower failure risk, reinforcing the role of sustained effort and conceptual clarity in inferential reasoning. However, students from non-ECAeI programs, those who report following instructions, and those who graduated from high schools in Antioquia consistently show higher probabilities of failure—even with similar perception levels or exam durations. These findings suggest that behavioral compliance does not necessarily imply deep engagement, and that regional or curricular differences may shape learning trajectories in subtle but significant ways. Taken together, the results reveal a complex network of academic, cognitive, and contextual influences on performance in advanced statistical content.

\section{Discussion}\label{sec-discussion}
The results obtained allow for a critical interpretation of how academic performance in statistics cannot be understood in a global or homogeneous manner. When data are disaggregated by course unit, differential patterns emerge that highlight the need for more nuanced and adaptive pedagogical approaches. One of the most consistent findings was the structural effect associated with students' academic programs: those enrolled in programs outside the School of Applied Sciences and Engineering (ECAeI)—particularly those in Interactive Design—consistently showed higher odds of failure across all four units of the course. This trend suggests a possible misalignment between the course’s instructional approach and the prior competencies or learning trajectories of these students, reinforcing the need to design differentiated strategies that ensure equitable learning conditions.

Another relevant factor was the academic stage at which the course was taken. Students who enrolled in the course during the early semesters of their programs had a higher probability of failing, especially in the units with greater conceptual complexity, such as Statistical Inference and Linear Regression. This finding suggests that academic maturity and the progressive development of abstract thinking skills may act as protective factors against the challenges of statistical reasoning.

Regarding perceived preparedness, its impact was evident but not consistent. In some units—particularly those involving more abstract content, such as hypothesis testing in regression or understanding inference exercises for sampling distributions, especially concerning the sample variance—a higher self-perception of competence was significantly associated with a lower probability of failure. However, in other units, this relationship was either non-existent or statistically weak, indicating that self-perception does not operate uniformly and that its predictive value may depend on the nature of the content assessed or the exam design.

Some variables also showed diminishing influence over the course duration. For example, the section taught by the full-time faculty member had a significant association with failure in Unit 1, but this difference disappeared in later units. This phenomenon may reflect a gradual adaptation by students to the instructor’s teaching style or an adjustment in evaluation criteria that resulted in more standardized pedagogical conditions.

Finally, several unexpected findings invite a reexamination of common assumptions in educational measurement. For instance, the higher risk of failure among students who evidenced following instructions may be due to coding issues, but it could also reflect superficial compliance behaviors that do not translate into meaningful content engagement. Similarly, the lower risk of failure among students who completed high school outside of Antioquia suggests that sociocultural or contextual factors may be influencing performance more deeply than previously measured.

\section{Conclusions \& Recommendations}\label{sec-conc}

The analysis confirms that failure in introductory statistics is not a homogeneous phenomenon but varies significantly across the different thematic units of the course. This finding supports the critique raised in the literature \cite{garfield2007how} regarding the need to move beyond global approaches and to promote instructional practices that distinguish between different forms of statistical reasoning. In particular, the Inference unit stands out as a critical barrier in the learning process, as documented in prior studies, due to the high cognitive demand involved in working with abstract concepts such as sampling distributions, confidence intervals, and hypothesis testing.

An additional contribution of this study is the empirical evidence on the relationship between self-perception and performance, which becomes clearer in units requiring greater conceptual reasoning. This result aligns with the arguments of \cite{bandura1997self} and \cite{finney2003self}, who emphasize that self-efficacy directly influences how students approach complex tasks. However, the study also revealed cases where self-perception had no predictive value, suggesting that subjective judgments of competence only exert a protective effect when accompanied by well-developed cognitive skills.

Another significant finding that emerged from the models is the impact of structural and contextual variables, such as academic program and stage of study. Students from non-ECAeI programs faced a consistent disadvantage across all units, which may be partially explained by a lack of alignment between the course’s focus and the students’ prior disciplinary training. This finding highlights the tension between maintaining common academic standards and adapting instruction to diverse student populations. Accordingly, there is a need to revise the curricular alignment of the General Statistics course with the programs that require it, particularly those that are not technically or quantitatively oriented.

From a pedagogical perspective, the results support the implementation of diagnostic assessments at the beginning of the course to identify vulnerable students early on, as well as differentiated support strategies tailored to each thematic unit. The design of targeted interventions—such as structured peer tutoring, contextualized learning materials, and activities that foster metacognitive reflection—may help bridge understanding gaps and reduce failure rates, especially in units like Inference and Regression. These efforts should be accompanied by careful monitoring to evaluate their impact through quasi-experimental research designs.

Finally, this study opens several lines for future research, including longitudinal tracking of students’ statistical skills throughout their academic programs, the integration of objective measures of prior academic performance, and a deeper qualitative exploration of the beliefs, emotions, and attitudes that shape students’ experiences with statistics. These approaches would enrich the understanding of the phenomenon, advance a more comprehensive view of learning in quantitative contexts, and inform educational policies that combine academic rigor with inclusion and equity.

\section*{Declarations}

\textbf{AI usage statement:}
\begin{quote}
	During the preparation of this work, I used ChatGPT (OpenAI, free version) to enhance readability and language. After employing this tool, I reviewed and edited the content as needed and take full responsibility for the content of the publication.
\end{quote}

\section{Disclosure statement}\label{disclosure-statement}

I declare that I have no conflicts of interest related to the research, authorship, or publication of this article.

\section{Data Availability Statement}\label{data-availability-statement}

Deidentified data have been made available at the following URL: \url{https://repository.eafit.edu.co/entities/publication/776b49b5-d51d-4ea5-8c14-b40c3d3fbe81}

\phantomsection\label{supplementary-material}
\bigskip

%
%
%

%

  \bibliography{bibliography.bib}

\end{document}